\documentclass[preprint,showpacs,preprintnumbers,amsmath,amssymb,endfloats*]{revtex4}
\usepackage{graphicx}

\begin{document}
\def \ee {\varepsilon}
\thispagestyle{empty}
\title{Lifshitz theory of atom-wall interaction with applications
to quantum reflection}

\author{V.~B.~Bezerra,${}^{1}$
G.~L.~Klimchitskaya,${}^{2,}$\footnote{on leave from
North-West Technical University, St.Petersburg, Russia}
V.~M.~Mostepanenko,${}^{2,}$\footnote{on leave from Noncommercial Partnership
``Scientific Instruments'',  Moscow,  Russia}
and C.~Romero${}^{1}$
}

\affiliation{
${}^{1}$Department of Physics, Federal University of Para\'{\i}ba,
C.P.5008, CEP 58059--970, Jo\~{a}o Pessoa, Pb-Brazil \\
${}^{2}$Center of Theoretical Studies and Institute for Theoretical
Physics, Leipzig University, Postfach 100920,
D-04009, Leipzig, Germany
}
\begin{abstract}
The Casimir-Polder interaction of an atom with a metallic wall is
investigated in the framework of the Lifshitz theory. It is
demonstrated that in some temperature (separation) region the
Casimir-Polder entropy takes negative values and goes to zero when
the temperature vanishes. This result is obtained both for an
ideal metal wall and for real metal walls. Simple analytical
representations for the Casimir-Polder free energy and force are
also obtained. These results are used to make a comparison between
the phenomenological potential used in the theoretical description
of quantum reflection and exact atom-wall interaction energy, as
given by the Lifshitz theory. Computations are performed for the
atom of metastable He${}^{\ast}$ interacting with metal (Au) and
dielectric (Si) walls. It is shown that the relative differences
between the exact and phenomenological interaction energies are
smaller in the case of a metallic wall. This is explained by the
effect of negative entropy which occurs only for a metal wall.
More accurate atom-wall interaction energies computed here can be
used for the interpretation of measurement data in the experiments
on quantum reflection.
\end{abstract}
\maketitle

\section{Introduction}

Atom-wall interaction, which is a special case of dispersion
interactions caused by quantum fluctuations, plays an important
role in different physical, chemical and biological phenomena
\cite{1}. At short separations from several angstroms to a few
nanometers the interaction of an atom with a wall is described by
the nonretarded van der Waals potential of the form $V_3=-C_3/a^3$
\cite{2}. At separations of about $1\,\mu$m the interaction
between an atom and a wall is of relativistic character and is
described by the Casimir-Polder potential $V_4=-C_4/a^4$ \cite{3}.
The Lifshitz theory \cite{4,5} provides a complete description of
atom-wall interaction including the transition region from the
nonrelativistic to relativistic regime. It also predicts that at
separations larger than a few micrometers thermal effects become
dominating. The Lifshitz theory takes into account realistic
properties of an atom and of a wall material. In so doing atom is
described by the dynamic polarizability as a function of frequency
and wall by the frequency-dependent dielectric permittivity. Over
a period of years many experiments were performed on measuring
atom-wall interaction in different separation regions. From the
theoretical side, more accurate expressions for the atom-wall
interaction energy were obtained (see Refs.~\cite{6,7,8} for the
literature on the subject and calculational results).

On the modern stage of experimental research of atom-wall
interaction, the magnitude and the distance dependence of the
Casimir-Polder force were confirmed in Ref.~\cite{9} when studying
the deflection of ground-state Na atoms passing through a
micron-sized parallel plate cavity. The experimental data were
compared with the theoretical position-dependent potential for an
atom between parallel ideal metal plates \cite{10,11}. Of special
interest are situations when the wave nature of atom becomes
dominant with respect to its classical behavior as a particle.
Such a pure quantum effect is what is called {\it quantum
reflection}, i.e., a process in which a particle moving through a
classically allowed region is reflected by a potential without
reaching a classical turning point. This is in fact a 
reflection of an ultracold atom under the influence of an {\it
attractive} atom-wall interaction or, in other words, the
above-barrier reflection. The phenomenon of quantum reflection has
become observable due to the success in the production of
ultracold atoms. First it was demonstrated with liquid surfaces
\cite{12,13,14,15} and then with Si, glass \cite{16},
$\alpha$-quartz crystal \cite{17} and Cu surfaces \cite{18}. It
was shown \cite{19} that quantum reflection is particularly
sensitive to details of the atom-surface interaction energy. On
the other hand,
 low-energy scattering from asymptotic power law
potentials on metallic surfaces was examined in Ref.~\cite{20}.

For theoretical calculation of the reflection amplitude, a simple
phenomenological atom-wall interaction potential is commonly used
\cite{17,21,22}, which is an interpolation between the nonretarded
van der Waals potential $V_3$ and the Casimir-Polder potential
$V_4$. The comparison of computational results with the
measurement data for the reflection amplitudes allows one to
estimate the parameters of the phenomenological potential.
However, the increasing precision of the measurements requests
comparison with the more accurate atom-wall interaction potentials
obtained from the Lifshitz theory and taking into account
realistic properties of the atom and the material of the wall.

In this paper we investigate the characteristic properties of the
interaction of atoms with metallic walls in the framework of the
Lifshitz theory. We start with the case of an ideal metal wall and
obtain the analytical expressions for the Casimir-Polder free
energy and force in different separation and temperature regions.
The delicate point that in the limit of short separation distances
the interaction energy depends on the velocity of light is
discussed. The asymptotic behavior of the Casimir-Polder entropy
at low temperatures (short separations), which is typical for any
model of the metal, is investigated. Negative values of the
entropy arising within some region of separations are linked to a
similar situation in the configuration of metal-dielectric
\cite{23,25,26}. Then, we summarize the calculation results for an
atom interacting with a wall made of real metal and obtain simple
analytical expressions for the Casimir-Polder interaction energy
and force applicable to moderate atom-wall separations. Special
attention is paid to the asymptotic behavior of the Casimir-Polder
entropy at low temperatures. It is shown that the entropy goes to
zero when the temperature vanishes, in accordance with the Nernst
heat theorem. (Recall that violation of Nernst's theorem occurs
for two parallel plates described by the Drude model with
vanishing relaxation parameter when temperature vanishes
\cite{27}; this situation was debated in the literature, see,
e.g., Refs.~\cite{28,29,28a,29a}.) However, we show that at low
temperatures the Casimir entropy of the atom-wall interaction for
real metal wall approaches zero from the negative values like this
occurs for an ideal metal wall. Finally we compare the
phenomenological potentials used in the calculation of reflection
amplitudes in quantum reflection with the accurate results of
numerical calculations using the Lifshitz formula. Both the cases
of metallic (Au) and dielectric (Si) walls interacting with an
atom of metastable He${}^{\ast}$ are considered. It is shown that
the maximum deviation between the phenomenological and accurate
results achieves 12.5\% for Si at a separation $a=300\,$nm and
10.5\% for Au at $a=400\,$nm. Increasing the separation, larger
deviations between the accurate and phenomenological potential
occur, especially for dielectric walls. It is shown that the
effect of negative Casimir-Polder entropy, discovered in this
paper in the configuration of an atom near a metal wall, makes the
deviation between the accurate and phenomenological potentials
smaller. Thus, with increasing experimental precision the use of
more accurate expressions for the interaction energy in the
calculation of reflection amplitudes may become preferable.

The paper is organized as follows. In Sec.~II we briefly summarize
the main results and notations used in the subsequent sections.
Section III contains consideration of an atom near an ideal metal
wall. In Sec.~IV simple analytical expressions for the case of
real metal wall are obtained. In Sec.~V the low-temperature
asymptotic behavior of the Casimir-Polder entropy is discussed.
Section VI is devoted to the comparison between the
phenomenological interaction energy commonly used to describe
quantum reflection and a more accurate one obtained from the
Lifshitz theory. Our conclusions and discussion are contained in
Sec.~VII.

\section{Lifshitz formulas for an atom near a wall}

In accordance with the Lifshitz theory, the free energy of an atom-wall
interaction for a wall at temperature $T$ in thermal equilibrium with
environment is given by
\begin{eqnarray}
&&
{\cal F}(a,T)=-k_BT
\sum_{l=0}^{\infty}{\vphantom{\sum}}^{\prime}
\alpha({\rm i}\xi_l)\int_{0}^{\infty}\!k_{\bot}dk_{\bot}
q_l{\rm e}^{-2aq_l}
\label{X6e15}\\
&&\times\left\{2r_{\rm TM}({\rm i}\xi_l,k_{\bot})
-\frac{\xi_l^2}{q_l^2c^2}\left[
r_{\rm TM}({\rm i}\xi_l,k_{\bot})+
r_{\rm TE}({\rm i}\xi_l,k_{\bot})
\right]\right\}\!.
\nonumber
\end{eqnarray}
\noindent
Here, $a$ is the separation distance between an atom and a wall,
$\alpha(\omega)$ is the atomic dynamic polarizability,
$\xi_l=2\pi k_BTl/\hbar$ with $l=0,\,1,\,2,\,\ldots$ are the Matsubara
frequencies, $k_B$ is the Boltzmann constant, $k_{\bot}$ is the projection
of the wave vector on the plane of a wall (i.e., perpendicular to the $z$
direction), prime near the summation sign means that a multiple 1/2
is added to the term with $l=0$, and $q_l^2=k_{\bot}^2+\xi_l^2/c^2$.
The reflection coefficients for two independent polarizations of the
electromagnetic field, transverse magnetic (TM) and transverse
electric (TE), are given by
\begin{equation}
r_{\rm TM}({\rm i}\xi_l,k_{\bot})=\frac{\varepsilon_lq_l-
k_l}{\varepsilon_lq_l+k_l}, \quad
r_{\rm TE}({\rm i}\xi_l,k_{\bot})=\frac{q_l-k_l}{q_l+k_l},
\label{X2e66}
\end{equation}
\noindent where $k_l^2=k_{\bot}^2+\varepsilon_l{\xi_l^2}/{c^2}$
and $\varepsilon_l=\varepsilon({\rm i}\xi_l)$ is the
frequency-dependent dielectric permittivity of the wall material
calculated at the imaginary Matsubara frequencies.

{}From Eq.~(\ref{X6e15}),
we obtain the Lifshitz-type formula for the Casimir-Polder force
acting on an atom  near a wall:
\begin{eqnarray}
&&
F(a,T)=-\frac{\partial{\cal F}(a,T)}{\partial a}
\nonumber \\
&&
=-2k_BT\sum_{l=0}^{\infty}{\vphantom{\sum}}^{\prime}
\alpha({\rm i}\xi_l)\int_{0}^{\infty}\!k_{\bot}dk_{\bot}
q_l^2{\rm e}^{-2aq_l}
\label{X6e16}\\
&&\times\left\{2r_{\rm TM}({\rm i}\xi_l,k_{\bot})
-\frac{\xi_l^2}{q_l^2c^2}\left[
r_{\rm TM}({\rm i}\xi_l,k_{\bot})+
r_{\rm TE}({\rm i}\xi_l,k_{\bot})
\right]\right\}\!.
\nonumber
\end{eqnarray}
\noindent The interesting characteristic feature of
Eqs.~(\ref{X6e15}) and (\ref{X6e16}) describing atom-wall
interaction is that the transverse electric reflection coefficient
at zero frequency, $r_{\rm TE}(0,k_{\bot})$, does not contribute
to the result as it is multiplied by  the factor $\xi_0^2=0$.
Because of this, in the case of metal wall the obtained values of
the free energy and force do not depend on the model used for the
metal. However, in the case of a dielectric wall, the  results
obtained depend on the transverse magnetic reflection coefficient
at zero frequency, $r_{\rm TM}(0,k_{\bot})$. This can be used as a
test for the model of the dielectric permittivity of the wall
material \cite{25,26,24,30,31,32,32KM}.

Below we perform some analytical calculations of the
van der Waals and Casimir-Polder interactions.
Also numerical computations are performed
for metastable He${}^{\ast}$ atoms and walls
made of different materials. For both these purposes it is useful to
express Eqs.~(\ref{X6e15}) and (\ref{X6e16}) in terms of the dimensionless
variables
\begin{eqnarray}
&&
y=2q_la, \qquad \zeta_l=\frac{\xi_l}{\omega_c}=\tau l,
\nonumber \\
&&
\tau=4\pi\frac{k_BaT}{\hbar c}=2\pi\frac{T}{T_{\rm eff}},
\label{X2e85}
\end{eqnarray}
\noindent where $\omega_c\equiv c/(2a)$ is the so-called {\it
characteristic frequency} and $T_{\rm eff}=\hbar\omega_c/k_B$ is
the {\it characteristic temperature}. Then, the Casimir-Polder
free energy for an atom near a wall is given by
\begin{eqnarray}
&&
{\cal F}(a,T)=-\frac{k_BT}{8a^3}
\sum_{l=0}^{\infty}{\vphantom{\sum}}^{\prime}
\alpha({\rm i}\zeta_l\omega_c)\int_{\zeta_l}^{\infty}\!dy
{\rm e}^{-y}
\label{X6e17}\\
&&\times\left\{2y^2r_{\rm TM}({\rm i}\zeta_l,y)
-\zeta_l^2\left[
r_{\rm TM}({\rm i}\zeta_l,y)+
r_{\rm TE}({\rm i}\zeta_l,y)
\right]\right\}.
\nonumber
\end{eqnarray}
\noindent
The reflection coefficients are expressed in these
variables in the following way:
\begin{eqnarray}
&&
r_{\rm TM}({\rm i}\zeta,y)=\frac{\varepsilon y-
\sqrt{y^2+\zeta^2(\varepsilon -1)}}{\varepsilon y+
\sqrt{y^2+\zeta^2(\varepsilon -1)}},
\label{X2e87} \\
&&
r_{\rm TE}({\rm i}\zeta,y)=\frac{y-\sqrt{y^2+\zeta^2(\varepsilon -1)}}{y+
\sqrt{y^2+\zeta^2(\varepsilon -1)}},
\qquad \varepsilon\equiv\varepsilon({\rm i}\omega_c\zeta).
\nonumber
\end{eqnarray}
\noindent
The respective expression for the Casimir force
acting on an atom is
\begin{eqnarray}
&&
{F}(a,T)=-\frac{k_BT}{8a^4}
\sum_{l=0}^{\infty}{\vphantom{\sum}}^{\prime}
\alpha({\rm i}\zeta_l\omega_c)\int_{\zeta_l}^{\infty}\!dy
y{\rm e}^{-y}
\label{X6e18}\\
&&~\times\left\{2y^2r_{\rm TM}({\rm i}\zeta_l,y)
-\zeta_l^2\left[
r_{\rm TM}({\rm i}\zeta_l,y)+
r_{\rm TE}({\rm i}\zeta_l,y)
\right]\right\}.
\nonumber
\end{eqnarray}
Note that here we consider atoms in the ground state and assume
that thermal radiation is not so strong to excite electrons to
higher states. The case of excited atoms is discussed in
Ref.~\cite{33}. The possible impact of virtual photon absorption
on the Casimir-Polder force was considered in Ref.~\cite{34}.

The Casimir-Polder free energy, as given in
Eq.~(\ref{X6e17}), can be represented in the form
\begin{equation}
{\cal F}(a,T)={\cal E}(a,T)+
\Delta{\mathcal F}(a,T).
\label{X6e19}
\end{equation}
\noindent
Here, the quantities ${\cal E}(a,T)$ and
$\Delta{\cal F}(a,T)$ are obtained from Eq.~(\ref{X6e17})
by the application of the Abel-Plana formula \cite{35,36}
\begin{equation}
\sum\limits_{n=0}^{\infty}
{\vphantom{\sum}}^{\prime} F(n)
=\int_{0}^{\infty}F(t)dt
+{\rm i}\int_{0}^{\infty}\frac{dt}{{\rm e}^{2\pi t}-1}
\left[F({\rm i}t)-F({-\rm i}t)\right].
\label{2e21}
\end{equation}
\noindent
They are given by
\begin{eqnarray}
&&
{\cal E}(a,T)=\frac{\hbar c}{32\pi a^4}\int_{0}^{\infty}\!\! d\zeta
\int_{\zeta}^{\infty}\!\!dy\,h(\zeta,y),
\label{X6e20} \\
&&h(\zeta,y)=-\alpha({\rm i}\omega_c\zeta){\rm e}^{-y}\left\{2y^2
r_{\rm TM}({\rm i}\zeta,y)\right.
\nonumber \\
&&~~~~~~~\left.
-\zeta^2\left[
r_{\rm TM}({\rm i}\zeta,y)+r_{\rm TE}({\rm i}\zeta,y)
\right]\right\},
\nonumber \\
&&
\Delta{\cal F}(a,T)=\frac{{\rm i}\hbar c\tau}{32\pi a^4}
\int_{0}^{\infty}\!\! dt\frac{H({\rm i}t\tau)-
H(-{\rm i}t\tau)}{{\rm e}^{2\pi t}-1},
\nonumber \\
&&
H(x)\equiv\int_{x}^{\infty}\!\!dy\,h(x,y).
\nonumber
\end{eqnarray}
\noindent
For the temperature-independent permittivities
one gets
${\cal E}(a,T)=E(a)$, where $E(a)$ is
the Casimir-Polder energy at zero
temperature, and $\Delta{\cal F}(a,T)=\Delta_T{\cal F}(a,T)$,
the thermal correction to it.

Equations (\ref{X6e15}) and (\ref{X6e16}) present the
Casimir-Polder energy and force at any separation between an atom
and a wall. In the nonrelativistic limit (short separations if
compared with the characteristic absorption wavelength of plate
material) summation in these equations can be replaced by
integration along continuous frequencies. At such short
separations the velocity of light $c$ can effectively be put equal
to infinity. Then, the Casimir-Polder (van der Waals) atom-wall
interaction energy is given by
\begin{eqnarray}
&&
E(a)=-\frac{\hbar}{2\pi}\int_{0}^{\infty}\!\!d\xi
\alpha({\rm i}\xi)\int_{0}^{\infty}\!\! k_{\bot}dk_{\bot}
\sqrt{k_{\bot}^2+\frac{\xi^2}{c^2}}\,
\nonumber  \\
&&~~~~~
\times {\rm e}^{-2a\sqrt{k_{\bot}^2+\frac{\xi^2}{c^2}}}
\left\{\vphantom{\frac{\xi^2}{c^2k_{\bot}^2+\xi^2}}
2r_{\rm TM}({\rm i}\xi,k_{\bot})\right.
\label{X6e21} \\
&&~~~~~\left.
-\frac{\xi^2}{c^2k_{\bot}^2+\xi^2}\left[
r_{\rm TM}({\rm i}\xi,k_{\bot})+
r_{\rm TE}({\rm i}\xi,k_{\bot})
\right]\right\}
\nonumber \\
&&~~\approx
-\frac{\hbar}{\pi}\int_{0}^{\infty}\!\!d\xi
\alpha({\rm i}\xi)\int_{0}^{\infty}\!\! k_{\bot}^2dk_{\bot}
{\rm e}^{-2ak_{\bot}}r_{\rm TM}({\rm i}\xi,k_{\bot}).
\nonumber
\end{eqnarray}
\noindent
In the nonrelativistic limit, from Eq.~(\ref{X2e66}) one
obtains
\begin{equation}
r_{\rm TM}({\rm i}\xi,k_{\bot})\approx
\frac{\varepsilon({\rm i}\xi)-1}{\varepsilon({\rm i}\xi)+1}.
\label{X6e22}
\end{equation}
\noindent
Substituting this result into Eq.~(\ref{X6e21}) one
arrives at the van der Waals energy for atom-wall interaction
\begin{equation}
V_3=E(a)=-\frac{C_3}{a^3}, \quad
C_3=\frac{\hbar}{4\pi }\int_{0}^{\infty}\!\!d\xi
\alpha({\rm i}\xi)
\frac{\varepsilon({\rm i}\xi)-1}{\varepsilon({\rm i}\xi)+1}.
\label{X6e23}
\end{equation}
\noindent
In the case of the relativistic limit of atom-wall
interaction, it is considered below separately for ideal and real
metals.

\section{Atom near an ideal metal plane}

The configuration of
an atom near metallic wall suggests some interesting and unexpected
features connected with the behavior of the Casimir-Polder entropy
at low temperature. This can be observed already in the simplest case
of an atom interacting with an ideal metal plane.
The Casimir-Polder free energy of atom-wall interaction is given by
 Eq.~(\ref{X6e15}).
{}From substitution of the reflection coefficients for an ideal
metal plane,
\begin{equation}
r_{\rm TM}({\rm i}\xi_l, k_{\bot})=1, \qquad
r_{\rm TE}({\rm i}\xi_l, k_{\bot})=-1,
\label{X4e1}
\end{equation}
\noindent
one obtains
\begin{equation}
{\cal F}(a,T)=-2k_BT\sum_{l=0}^{\infty}{\vphantom{\sum}}^{\prime}
\alpha({\rm i}\xi_l)\int_{0}^{\infty}\!\!k_{\bot}dk_{\bot}q_l
{\rm e}^{-2aq_l}.
\label{X6e24}
\end{equation}
\noindent
In the region of short and moderate separation distances between an atom
and a metal plane, where the thermal effects can be neglected, the free
energy is approximately equal to the energy. The latter can be obtained
from  Eqs.~(\ref{X6e21}) and (\ref{X4e1})
\begin{equation}
E(a)=-\frac{\hbar}{\pi}\int_{0}^{\infty}\!\!d\xi
\alpha({\rm i}\xi)\int_{0}^{\infty}\!\!k_{\bot}dk_{\bot}q
{\rm e}^{-2aq}.
\label{X6e25}
\end{equation}
\noindent
Introducing the dimensionless variables $y=2aq$,
$\zeta=\xi/\omega_c$  and integrating with respect to $y$, we can
rearrange this equation to the form
\begin{equation}
E(a)=-\frac{\hbar c}{16\pi a^4}\int_{0}^{\infty}\!\!d\zeta
\alpha({\rm i}\omega_c\zeta)(\zeta^2+2\zeta+2)
{\rm e}^{-\zeta}.
\label{X6e26}
\end{equation}

If we consider moderate separations from about $1\,\mu$m to
$3\,\mu$m only, the approximation of a static atomic
polarizability, $\alpha({\rm i}\omega_c\zeta)\approx\alpha(0)$,
works well. In this case, Eq.~(\ref{X6e26}) leads to
\begin{equation}
V_4=E(a)\equiv E_{\rm CP}(a) =-\frac{C_4}{a^4}=
-\frac{3\hbar c}{8\pi a^4}
\alpha(0).
\label{X6e27}
\end{equation}
\noindent
 This result was first obtained by Casimir and Polder
\cite{3} and corresponds to
 the relativistic atom-wall interaction potential.

At all separations larger than 50--70\,nm
the single-oscillator model for
the dynamic atomic polarizability leads to less than 1\% deviations
from the calculational results using highly
accurate atomic polarizabilities \cite{8}.
The single-oscillator model presents $\alpha({\rm i}\xi)$
in the form
\begin{equation}
\alpha({\rm i}\xi)=\frac{\alpha(0)}{1+\frac{\xi^2}{\omega_0^2}}=
\frac{\alpha(0)}{1+\frac{\omega_c^2}{\omega_0^2}\zeta^2}\equiv
\alpha({\rm i}\omega_c\zeta).
\label{X6e28}
\end{equation}
\noindent
 Here, $\omega_0$ is the characteristic absorption
frequency for the atom under consideration. We perform calculation
under the assumption that
$\beta_A\equiv\omega_c/\omega_0=\lambda_0/(4\pi a)\gg 1$, where
$\lambda_0=2\pi c/\omega_0$ is the characteristic absorption
wavelength of an atom. This assumption is satisfied at atom-plane
separations $a\ll\lambda_0$. We have once more introduced  the new
variables in Eq.~(\ref{X6e25}), but changed the order of
integrations in $\zeta$ and $y$ rather than first integrate with
respect to $y$ as was done previously. The result is
\begin{equation}
E(a)=-\frac{\hbar c}{16\pi a^4}\int_{0}^{\infty}\!\!y^2dy
{\rm e}^{-y}
\int_{0}^{y}\!\!d\zeta
\alpha({\rm i}\omega_c\zeta).
\label{X6e29}
\end{equation}
\noindent
Substituting Eq.~(\ref{X6e28}) into Eq.~(\ref{X6e29})
and integrating in $\zeta$,
we get
\begin{equation}
E(a)=-\frac{\hbar c}{16\pi a^4}\alpha(0)\frac{1}{\beta_A}
\int_{0}^{\infty}\!\!y^2dy
{\rm e}^{-y}
{\rm arctan}(\beta_Ay).
\label{X6e30}
\end{equation}
\noindent
Taking into account that $\beta_A\gg 1$, it is possible to replace
${\rm arctan}(\beta_Ay)$ with $\pi/2$ without loss of accuracy.
This leads to
\begin{equation}
E(a)=-\frac{\hbar c}{4\lambda_0 a^3}\alpha(0).
\label{X6e31}
\end{equation}
\noindent
 Note that although in Eq.~(\ref{X6e31}) the same
distance dependence  appears (inverse third power of separation),
as in the nonrelativistic limit in Eq.~(\ref{X6e23}), it is quite
different in nature. Particularly Eq.~(\ref{X6e23}) does not
contain the velocity of light, as is appropriate for the
nonrelativistic limit, whereas Eq.~(\ref{X6e31}) does. In fact,
the nonrelativistic limit cannot be achieved for an ideal metal
wall, because the characteristic absorption wavelength of an ideal
metal is zero. The dependence of the Casimir-Polder energy at
short separation distances in  Eq.~(\ref{X6e31}) on the velocity
of light  should be considered as an indication that at these
separations the approximation of ideal metal is not applicable.

Now, we consider any separation distance larger than $1\,\mu$m
including those larger than $3\,\mu$m. Here, one can use the
static atomic polarizability, but should take thermal effects into
account. Introducing the dimensionless variables (\ref{X2e85}) we
rearrange the free energy given by Eq.~(\ref{X6e24}) into the form
\begin{equation}
{\cal F}(a,T)=-\frac{k_BT}{4a^3}\alpha(0)
\sum_{l=0}^{\infty}{\vphantom{\sum}}^{\prime}
\int_{\zeta_l}^{\infty}y^2dy{\rm e}^{-y}.
\label{X6e32}
\end{equation}
\noindent
After performing the integration and summation one obtains
\begin{equation}
{\cal F}(a,T)=E_{\rm CP}(a)\,\eta(a,T),
\label{X6e33}
\end{equation}
\noindent
where the Casimir-Polder energy is given in Eq.~(\ref{X6e27})
and the correction factor is
\begin{equation}
\eta(a,T)=\frac{\tau}{6}\left[1+\frac{2}{{\rm e}^{\tau}-1}+
\frac{2\tau{\rm e}^{\tau}}{({\rm e}^{\tau}-1)^2}+
\frac{\tau^2{\rm e}^{\tau}({\rm e}^{\tau}+1)}{({\rm e}^{\tau}-1)^3}
\right].
\label{X6e34}
\end{equation}
\noindent
Note that the parameter $\tau$, defined
in Eq.~(\ref{X2e85}), is
linear in separation and temperature.
 The asymptotic behavior of the correction factor
(\ref{X6e34}) at low temperature is given by
\begin{equation}
\eta(a,T)=1-\frac{\tau^4}{2160}+\frac{\tau^6}{15120}-
\frac{\tau^8}{241920}+O(\tau^{10}).
\label{X6e35}
\end{equation}

The Casimir-Polder force between an atom and an ideal metal plane
can also be presented in a form similar to  Eq.~(\ref{X6e33}). For
this purpose one can use Eq.~(\ref{X6e18}) with the reflection
coefficients (\ref{X4e1}) or calculate the negative derivative
with respect to $a$ from both sides of Eq.~(\ref{X6e33}). The
result is
\begin{equation}
F(a,T)=F_{\rm CP}(a)\kappa(a,T), \quad
F_{\rm CP}(a)=-\frac{3\hbar c}{2\pi a^5}\alpha(0).
\label{X6e36}
\end{equation}
\noindent
The correction factor to the Casimir-Polder force $F_{\rm CP}(a)$
can be represented as
\begin{equation}
\kappa(a,T)=\frac{3}{4}\eta(a,T)+
\frac{\tau^4{\rm e}^{\tau}({\rm e}^{2\tau}+
4{\rm e}^{\tau}+1)}{24({\rm e}^{\tau}-1)^4}.
\label{X6e37}
\end{equation}
\noindent
 The asymptotic behavior of this correction factor
 at low temperature is given by
\begin{equation}
\kappa(a,T)=1-\frac{\tau^6}{30240}-
\frac{\tau^8}{241920}+O(\tau^{10}).
\label{X6e38}
\end{equation}

It is interesting that representations like (\ref{X6e33}) and
(\ref{X6e35}) for the free energy or (\ref{X6e36}) and (\ref{X6e38})
for the force of atom-wall interaction can be obtained directly from the
Lifshitz formulas describing the Casimir interaction between
metallic and dielectric plates when the dielectric plate is
dilute \cite{23,25,26}.

Now, we are in a position to find the entropy of the
Casimir-Polder interaction. Calculating the negative derivative of
Eq.~(\ref{X6e33}) with respect to the temperature, we get the
expression
\begin{equation}
 S(a,T)=\frac{3k_B}{2a^3}\alpha(0)\sigma(a,T),
 \label{X6e42}
 \end{equation}
 \noindent
 where
 \begin{equation}
\sigma(a,T)=\frac{1}{\tau}\eta(a,T)- \frac{\tau^3{\rm
e}^{\tau}({\rm e}^{2\tau}+ 4{\rm e}^{\tau}+1)}{6({\rm
e}^{\tau}-1)^4}.
\label{X6e42a}
\end{equation}
\noindent
 It can easily be seen that the asymptotic expansion of the entropy
 factor,
$\sigma$, at low temperature is given by
\begin{equation}
\sigma(a,T)=-\frac{\tau^3}{540}+
\frac{\tau^5}{2520}+O(\tau^7).
\label{X6e43}
\end{equation}
\noindent Thus, the Casimir-Polder entropy goes to zero when
temperature vanishes in accordance with the Nernst heat theorem.
Note, however, that at low temperatures (small $\tau$) the entropy
(\ref{X6e42}) takes negative values. In Fig.\ 1 we plot the
entropy factor $\sigma$ from Eq.~(\ref{X6e42}) in the
configuration of an atom near an ideal metal plane as a function
of $\tau$. As is seen in Fig.\ 1, the Casimir-Polder entropy is
negative for $0<\tau<3$ and positive for larger $\tau$. This is in
accordance with the respective results for the configuration of
metal and dielectric plates \cite{23,25,26}. Keeping in mind that
the Lifshitz formula for an atom near a metal plate  is obtained
from the formula describing the two parallel plates one of which
being metallic and the other a dilute dielectric, the  similarity
obtained in the behavior of entropy appears quite natural.

In the high temperature limit $T\gg T_{\rm eff}$, only the
zero-frequency term in Eq.~(\ref{X6e32}) determines the total result,
whereas all terms with $l\geq 1$ are exponentially small. In this
case Eq.~(\ref{X6e32}) leads to
\begin{equation}
{\cal F}(a,T)=-\frac{k_BT}{4a^3}\alpha(0).
\label{X6e44}
\end{equation}
\noindent
This is the classical limit \cite{37,38} of the
Casimir-Polder free energy because the right-hand side of
Eq.~(\ref{X6e44}) does not depend on $\hbar$. In this situation,
the respective expressions for the Casimir-Polder entropy and
force are given by
\begin{equation}
S(a,T)=-\frac{k_B}{4a^3}\alpha(0), \qquad
F(a,T)=-\frac{3k_BT}{4a^4}\alpha(0).
\label{X6e45}
\end{equation}

\section{Atom near a real metal plate}

Here, we consider a metal plate made of Au described by the plasma
model
\begin{equation}
\varepsilon(\omega)=1-\frac{\omega_p^2}{\omega^2}
\label{X3e1}
\end{equation}
\noindent
 with the plasma frequency $\omega_p=9.0\,$eV.
This allows rather precise results at separation distances larger
than the plasma wavelength $\lambda_p=137\,$nm. At these
separations the dynamic polarizability of an atom can be
represented using the single-oscillator model (\ref{X6e28}). For
example, for the metastable helium atom He${}^{\ast}$ we have
$\alpha(0)=315.63\,$a.u. (one a.u. of polarizability is equal to
$1.482\times 10^{-31}\,\mbox{m}^{3}$),
$\omega_0=1.18\,\mbox{eV}=1.794\times 10^{15}\,$rad/s \cite{39}.
Equation (\ref{X6e28}) with the above value of $\omega_0$ is
appropriate in the frequency region contributing to the
Casimir-Polder interaction. This was demonstrated in Ref.~\cite{8}
by comparing the computational results obtained using the
single-oscillator model with that using the highly accurate atomic
dynamic polarizability.

Under these conditions the correction factors $\eta(a,T)$ and
$\kappa(a,T)$ to the Casimir-Polder free energy and force,
respectively, were computed in Ref.~\cite{6} using the Lifshitz
formulas (\ref{X6e17}) and (\ref{X6e18}). It was shown that at
short separations the effect of the nonzero skin depth of the
metal wall is much greater for an atom described by the static
polarizability than for an atom described by its dynamic
polarizability. In particular, for a real atom characterized by
the dynamic polarizability the corrections due to nonzero skin
depth of a metal wall are much less than for two metal plates. For
example, for two parallel plates the use of the plasma model
instead of the tabulated optical data at the separations
considered leads to an error of less than 2\% in the free energy
\cite{36}. For atom-wall interaction, however, the use of the
plasma model leads to less than 1\% error in the values of the
Casimir-Polder free energy and force compared to the use of
$\varepsilon({\rm i}\xi)$ obtained from the complete tabulated
optical data.  One can also conclude \cite{6} that at shorter
separations a proper  account of the dynamic atomic polarizability
is more important than that of the nonzero skin depth. At
intermediate separation distances from 1 to $3\,\mu$m the dynamic
atomic polarizability and the nonzero skin depth of the metal play
qualitatively equal roles. With increasing $a$ the role of the
dynamic polarizability becomes negligible, and the free energy is
determined solely by $\alpha(0)$. The high-temperature asymptotic
expression (\ref{X6e44}) becomes applicable at $a>6\,\mu$m. The
overall conclusion is that corrections due to nonzero skin depth,
dynamic atomic polarizability and nonzero temperature should be
taken into account in precise esperiments. In the case of the
Casimir-Polder force, the correction factors play even a stronger
role than in the case of the free energy. In Sec.~VI numerical
computations using the Lifshitz formula (\ref{X6e17}) are
performed in a wide separation region. The dielectric permittivity
of wall material is found from the complex index of refraction
\cite{40}. The results are compared with the phenomenological
potentials used in the theoretical description of quantum
reflection.

Within the separation distance from 1 to $3\,\mu$m, where the thermal
correction is small, the role of corrections to the Casimir-Polder
energy due to the nonzero skin depth and dynamic polarizability can be
illustrated analytically. For this purpose one can start from the
plasma and single-oscillator models and use the perturbative expansion
in the relative skin depth $\delta_0/a=\lambda_p/(2\pi a)$
and in the oscillator parameter $\beta_A$. Expanding the function
$h(\zeta,y)$ in Eq.~(\ref{X6e20}) up to the second power in both
parameters, we obtain
\begin{eqnarray}
h(\zeta,y)&=&-\alpha(0){\rm e}^{-y}\left[
\vphantom{\left(\frac{\delta_0}{a}\right)^2}
2y^2-2\beta_A^2\zeta^2y^2+\left(\frac{\zeta^4}{y}-3\zeta^2 y\right)
\frac{\delta_0}{a}\right.
\nonumber \\
&+&\left.\frac{1}{2}\left(2\zeta^4-\frac{\zeta^6}{y^2}+\zeta^2 y^2\right)
\left(\frac{\delta_0}{a}\right)^2\right].
\label{X6e46}
\end{eqnarray}
\noindent
Now we substitute (\ref{X6e46}) into the first equality of Eq.~(\ref{X6e20}),
change the order of integrations and calculate integrals with respect to
$\zeta$ and to $y$. The result is
\begin{equation}
E(a)=E_{\rm CP}(a)\left[1-\frac{20}{3}\beta_A^2-\frac{8}{5}\,
\frac{\delta_0}{a}+\frac{62}{21}\left(\frac{\delta_0}{a}\right)^2\right],
\label{X6e47}
\end{equation}
\noindent
where $E_{\rm CP}(a)$ is defined in Eq.~(\ref{X6e27}).
Substituting the above parameters for the Au wall and He${}^{\ast}$ atom,
we find that at $a=1\,\mu$m the correction to unity due to nonzero skin
depth is equal to --0.034, whereas the correction due to dynamic
polarizability is equal to --0.046. At $a=2\,\mu$m these corrections are
--0.018 and --0.012, respectively, and they decrease further  with the
increase of separation. {}From Eq.~(\ref{X6e47}) it can be seen that at a
separation distance of about $1\,\mu$m the corrections to the Casimir
energy due to the nonzero skin depth and dynamic polarizability of the
atom play a qualitatively equal role, as was discussed on the basis of the
numerical computations.

The respective expression for the Casimir-Polder force is obtained
as being the negative derivative of both sides of
Eq.~(\ref{X6e47}) with respect to $a$
\begin{equation}
F(a)=F_{\rm CP}(a)\left[1-10\beta_A^2-2\frac{\delta_0}{a}
+\frac{31}{7}\left(\frac{\delta_0}{a}\right)^2\right],
\label{X6e47a}
\end{equation}
\noindent
where $F_{\rm CP}(a)$ is defined in Eq.~(\ref{X6e36}).

\section{Asymptotic behavior at low temperature}

We now turn our attention to the examination of the
low-temperature behavior of the Casimir-Polder free energy,
entropy and force for an atom interacting with a metallic wall
made of real metal. This allows one to solve complicated problems
on the consistency of the Lifshitz theory, as adapted for
atom-wall interaction, with thermodynamics. The asymptotic
expressions obtained in this section can also serve as a test for
some  generalizations of the Lifshitz theory. As above, we
describe a real metal by means of the plasma model and the atom
with a single-oscillator expression for the dynamic
polarizability. Thus, separation distances larger than 150\,nm are
applicable.

We start once again from the function $h(x,y)$ defined in
Eq.~(\ref{X6e20}) but expand it  to the second power of only the
parameter $\delta_0/a$
\begin{eqnarray}
&&h(x,y)=-\frac{\alpha(0)}{1+\beta_A^2x^2}\,{\rm e}^{-y}\left[
\vphantom{\left(\frac{\delta_0}{a}\right)^2}
2y^2+\left(\frac{x^4}{y}-3x^2y\right)\frac{\delta_0}{a}\right.
\nonumber \\
&&~~
\left.+\frac{1}{2}\left(2x^4-\frac{x^6}{y^2}+x^2y^2\right)
\left(\frac{\delta_0}{a}\right)^2\right]
\label{X6e48} \\
&&~~
\equiv
h_0(x,y)+h_1(x,y)+h_2(x,y),
\nonumber
\end{eqnarray}
\noindent
where $h_k(x,y)$ ($k=0,\,1,\,2$) are the contributions
to $h(x,y)$ of order $(\delta_0/a)^k$. The function $H(x)$ defined
in Eq.~(\ref{X6e20}) is given by
\begin{eqnarray}
&&
H(x)=H_1(x)+H_2(x)+H_3(x),
\label{X6e49} \\
&&
H_k(x)=\int_{x}^{\infty}\!\!dyh_k(x,y).
\nonumber
\end{eqnarray}
\noindent
To calculate the thermal correction to the
Casimir-Polder energy defined in Eq.~(\ref{X6e20}) one needs to
find the difference $H({\rm i}t\tau)-H(-{\rm i}t\tau)$. This is
most easily done for every $H_k(x)$ separately. Thus, for $k=0$
\begin{eqnarray}
&&
H_0(x)=-\frac{2\alpha(0)}{1+\beta_A^2x^2}\int_{x}^{\infty}\!\!
dy{\rm e}^{-y}y^2
\nonumber \\
&&\phantom{H_0(x)}
=-\frac{2\alpha(0)}{1+\beta_A^2x^2}{\rm e}^{-x}
(2+2x+x^2).
\label{X6e50}
\end{eqnarray}
\noindent
Expanding this in powers of $x$ we obtain
\begin{eqnarray}
&&
H_0({\rm i}t\tau)-H_0(-{\rm i}t\tau)=-4{\rm i}\alpha(0)\tau^3t^3\left[
\frac{1}{3}\vphantom{\left(\frac{\beta_A^2}{3}\right)}\right.
\label{X6e51} \\
&&~~~~\left.
-\left(\frac{1}{10}-\frac{\beta_A^2}{3}\right)\tau^2t^2+
\left(\frac{1}{168}-\frac{\beta_A^2}{10}+\frac{\beta_A^4}{3}\right)
\tau^4t^4\right].
\nonumber
\end{eqnarray}
\noindent
Substituting Eq.~(\ref{X6e51}) into the third equality
of Eq.~(\ref{X6e20}) and integrating in $t$, we find the
respective contribution to the thermal correction which is given
by
\begin{eqnarray}
\Delta_T{\mathcal F}_0(a,T)&=&\frac{\hbar c\alpha(0)}{128\pi a^4}\,
\tau^4\left[\frac{1}{45}-\frac{\tau^2}{315}
\left(1-\frac{10}{3}\beta_A^2\right)\right.
\nonumber \\
&+&\left.
\frac{\tau^4}{5040}
\left(1-\frac{84}{5}\beta_A^2+56\beta_A^4\right)\right].
\label{X6e52}
\end{eqnarray}
\noindent
For $\beta_A=0$ this is just the thermal correction for an atom described
by the static polarizability near an ideal metal plane calculated
using a different method
in Sec.~III and contained in Eqs.~(\ref{X6e33}) and (\ref{X6e35}).

In a similar way, for $k=1$ one has
\begin{eqnarray}
&&
H_1(x)=-\frac{\alpha(0)}{1+\beta_A^2x^2}\frac{\delta_0}{a}
\int_{x}^{\infty}\!\!dy{\rm e}^{-y}\left(\frac{x^4}{y}-3x^2y\right)
\nonumber \\
&&~~
=\frac{\alpha(0)}{1+\beta_A^2x^2}\frac{\delta_0}{a}
\left[3x^2{\rm e}^{-x}(1+x)-x^4\Gamma(0,x)\right],
\label{X6e53}
\end{eqnarray}
\noindent
where $\Gamma(z,x)$ is the incomplete gamma function.
Expanding this in powers of $x$, we obtain
\begin{eqnarray}
&&
H_1({\rm i}t\tau)-H_1(-{\rm i}t\tau)={\rm i}\alpha(0)\tau^4t^4
\frac{\delta_0}{a}
\label{X6e54}\\
&&~~~~~
\times
\left(\pi+\pi\beta_A^2\tau^2t^2-
\frac{4}{45}\tau^3t^3\right).
\nonumber
\end{eqnarray}
\noindent
The contribution from this to the
thermal correction is given by
\begin{eqnarray}
&&
\Delta_T{\cal F}_1(a,T)=-\frac{\hbar c\alpha(0)}{128\pi a^4}\,
\tau^5\frac{\delta_0}{a}
\label{X6e55} \\
&&~~~
\times
\left[\frac{3\zeta_R(5)}{\pi^4}
+\beta_A^2\tau^2\frac{45\zeta_R(7)}{2\pi^6}-
\frac{\tau^3}{1350}\right].
\nonumber
\end{eqnarray}
\noindent
For $k=2$ one arrives at
\begin{eqnarray}
&&
H_2(x)=-\frac{\alpha(0)}{2(1+\beta_A^2x^2)}\left(\frac{\delta_0}{a}\right)^2
\int_{x}^{\infty}\!\!dy{\rm e}^{-y}\left(2x^4-\frac{x^6}{y^2}+x^2y^2\right)
\label{X6e56} \\
&&~
=-\frac{\alpha(0)}{2(1+\beta_A^2x^2)}\left(\frac{\delta_0}{a}\right)^2
\left[2x^4{\rm e}^{-x}-x^6\Gamma(-1,x)+x^2{\rm e}^{-x}(2+2x+x^2)\right].
\nonumber
\end{eqnarray}
\noindent
After expanding in powers of $x$ the following equality
is obtained:
\begin{eqnarray}
&&
H_2({\rm i}t\tau)-H_2(-{\rm i}t\tau)=\frac{{\rm i}\alpha(0)}{2}\tau^5t^5
\left(\frac{\delta_0}{a}\right)^2
\label{X6e57} \\
&&~~
\times
\left[\frac{20}{3}-\pi\tau t+
\frac{2}{15}(1+50\beta_A^2)\tau^2t^2\right].
\nonumber
\end{eqnarray}
\noindent
This implies that respective contribution  to the
thermal correction is
\begin{eqnarray}
&&
\Delta_T{\cal F}_2(a,T)=-\frac{\hbar c\alpha(0)}{128\pi a^4}\,
\tau^6\left(\frac{\delta_0}{a}\right)^2
\label{X6e58} \\
&&~~
\times
\left[\frac{5}{189}-
\frac{45\zeta_R(7)}{4\pi^6}\tau+\frac{1}{1800}(1+50\beta_A^2)\tau^2
\right].
\nonumber
\end{eqnarray}

Taking together Eqs.~(\ref{X6e52}), (\ref{X6e55}) and (\ref{X6e58}), we
find the low-temperature asymptotic behavior of the Casimir-Polder free
energy
\begin{eqnarray}
&&
\Delta_T{\mathcal F}(a,T)=\frac{\hbar c\alpha(0)}{128\pi a^4}\,
\tau^4\left\{\frac{1}{45}-\frac{\tau^2}{315}
\left(1-\frac{10}{3}\beta_A^2\right)\right.
\label{X6e59} \\
&&~~~
+\frac{\tau^4}{5040}
\left(1-\frac{84}{5}\beta_A^2+56\beta_A^4\right)-
\tau\frac{\delta_0}{a} \left[\frac{3\zeta_R(5)}{\pi^4}
+\beta_A^2\tau^2\frac{45\zeta_R(7)}{2\pi^6}-
\frac{\tau^3}{1350}\right]
\nonumber \\
&&~~~
-\left.\tau^2\left(\frac{\delta_0}{a}\right)^2 \left[\frac{5}{189}-
\frac{45\zeta_R(7)}{4\pi^6}\tau+\frac{1}{1800}(1+50\beta_A^2)\tau^2
\right]\right\}.
\nonumber
\end{eqnarray}
\noindent
This expression includes the effect of both the nonzero skin depth of the
metal plate and the dynamic polarizability of the atom. Several terms on the
right-hand side of Eq.~(\ref{X6e59}) do not contribute to the Casimir-Polder
force because the quantities $\tau/a$ and $\tau\beta_A$ do not depend
on the separation distance $a$. Calculating the negative derivative of
Eq.~(\ref{X6e59}) with respect to $a$, one obtains the thermal correction
to the Casimir-Polder force at zero temperature (\ref{X6e47a})
\begin{eqnarray}
&&
\Delta_T{F}(a,T)=\frac{\hbar c\alpha(0)}{128\pi a^5}\,
\tau^6\left[\frac{2}{315}-\frac{\tau^2}{30}
\left(\frac{1}{42}-\frac{1}{5}\beta_A^2\right)\right.
\nonumber \\
&&~~~~
-3\tau^2\frac{\delta_0}{a}
-\left.\tau\left(\frac{\delta_0}{a}\right)^2 \left(
\frac{45\zeta_R(7)}{4\pi^6}-2\tau\right)
\right].
\label{X6e60}
\end{eqnarray}
\noindent
At $\delta_0=\beta_A=0$ this expression coincides with the thermal
correction contained in Eqs.~(\ref{X6e36}) and (\ref{X6e38}) derived
for an ideal metal wall interacting with an atom characterized by the
static polarizability.

Equation (\ref{X6e59}) allows the calculation of the Casimir-Polder
entropy at low temperature. Calculating the negative derivative with
respect to temperature of both sides of Eq.~(\ref{X6e59}), we arrive at
\begin{eqnarray}
&&
S(a,T)=-\frac{k_B\alpha(0)}{32a^3}\,
\tau^3\left[\frac{4}{45}-\frac{2\tau^2}{105}
\left(1-\frac{10}{3}\beta_A^2\right)\right.
\nonumber \\
&&~~~~
-\left.\tau\frac{\delta_0}{a}
\frac{15\zeta_R(5)}{\pi^4}
-\frac{10}{63}\tau^2\left(\frac{\delta_0}{a}\right)^2
\right].
\label{X6e61}
\end{eqnarray}
\noindent
As we can see, this entropy goes to zero when the
temperature vanishes, implying that the Nernst heat theorem is
satisfied. Although  in the derivation of Eq.~(\ref{X6e61}) we
used the plasma model, this conclusion is valid for any other
approach to the description of real metals, including the Drude
model approach \cite{29}. The point to note is that the TE
reflection coefficient at zero frequency does  not contribute to
the Casimir-Polder atom-wall interaction. Regarding the
contributions of all other Matsubara frequencies and the TM
reflection coefficient at $\xi=0$, different theoretical
approaches to the description of a real metal in the framework of
the Lifshitz theory lead to practically coincident results
\cite{27}. Thus the standard Lifshitz theory of atom-wall
interaction in the case of a metal wall is thermodynamically
consistent. At the same time, as seen from Eq.~(\ref{X6e61}),
$S(a,T)$ at low temperature is negative. Thus, this property of
the atom-wall configuration, discussed above in the case of an
ideal metal wall (Sec.~III), is also preserved for real metal
walls. Note that the asymptotic expressions (\ref{X6e44}),
(\ref{X6e45}) obtained for an atom near ideal metal wall at high
temperature are valid for real metal wall as well. Nonzero skin
depth does not play any role at high temperatures (large
separations).

In contrast to the case of metal wall, the interaction of the atom with
dielectric walls runs into problems when the dc conductivity of the wall
material is included into the model of the dielectric response.
Specifically, in this case the Casimir entropy calculated in the
framework of the Lifshitz theory takes positive value at zero
temperature, i.e., the Nernst theorem is violated \cite{41}.
Theoretical results for atom-wall interaction calculated with
included small dc conductivity of a dielectric wall are shown to be
experimentally inconsistent \cite{32KM}.

\section{Accuracy of the phenomenological potential used to describe
quantum reflection}

Practically all papers devoted to the investigation of quantum
reflection (see, e.g., Refs.~\cite{16,17,21,22,42,43,44,45}) use the
phenomenological potential of atom-wall interaction at zero
temperature to calculate the reflection amplitude. In fact,
the reflection amplitude for ultra cold atoms depends critically
on the shape of the potential in two asymptotic regions of small and
large atom-wall separations. Specifically, it was shown \cite{42}
that the reflection amplitude depends on the dimensionless parameter
\begin{equation}
\rho=\frac{\sqrt{2m_a}}{\hbar}\,\frac{C_3}{\sqrt{C_4}},
\label{eqR}
\end{equation}
\noindent 
where $m_a$ is the mass of an atom.
In this case for $\rho<1$ the nonretarded interaction (\ref{X6e23})
is dominant in the quantum reflection, whereas for $\rho>1$ the retarded
interaction (\ref{X6e27}) becomes dominant. For atoms of metastable
He${}^{\ast}$ interacting with Au or Si walls considered below the
parameter $\rho$ is large and the reflection amplitude mostly depends
on the Casimir-Polder potential \cite{21,42,43}. The shape of the
potential in the transition region between the nonretarded and retarded
interactions may contribute to the reflection amplitude markedly only
for atoms with increased velocity.
Below we compare the phenomenological potentials used in quantum
reflection with more accurate interaction energies computed using the
Lifshitz formula at both zero and nonzero temperature.

The most often used phenomenological potential
(interaction energy) has the form
\begin{equation}
E(a)=-\frac{C_4}{a^3(a+l)},
\label{eqn1}
\end{equation}
\noindent
where $l$ is a characteristic parameter with the
dimension of length that depends on the material. It is supposed
that at short separations $a\ll l$ (typically at separations of
order a few nanometers), $E(a)$ coincides with the van der Waals
potential $V_3$, so that $C_4=lC_3$. The coincidence between
$E(a)$  and the Casimir-Polder potential $V_4$ is achieved at
separations of about $10\,\mu$m where $l$ is negligibly small in
comparison with $a$. At such large separations the correction
factor to the Casimir-Polder energy due to nonzero skin depth and
dynamic atomic polarizability is practically equal to unity.

As an example we consider the Au wall and the atom of metastable He${}^{\ast}$.
Using the value of $\alpha(0)$ presented in Sec.~IV, one obtains from
Eq.~(\ref{X6e27}) the magnitude of the Casimir-Polder coefficient
$C_4^{\rm Au}\approx 1.1\,\mbox{eV\,nm}^4\approx
1.8\times 10^{-55}\,\mbox{J\,m}^4$. The value of the van der Waals
coefficient $C_3$ for Au can be computed from Eq.~(\ref{X6e23}).
In so doing one should use the tabulated optical data for the complex
index of refraction of Au \cite{40} in order to find the values of
$\varepsilon$ along the imaginary axis, and the highly accurate data
for the dynamic polarizability of He${}^{\ast}$ atom [at short separations
of a few nanometers the plasma model (\ref{X3e1}) and the single-oscillator
model (\ref{X6e28}) are not applicable in precise computations).
The computations (see Ref.~\cite{8} for details) lead to
$C_3^{\rm Au}\approx 1.6\,\mbox{a.u.}\approx
6.4\times 10^{-3}\,\mbox{eV\,nm}^4\approx 10.2\times 10^{-49}\,\mbox{J\,m}^4$.
{}From this we obtain
$l^{\rm Au}=C_4^{\rm Au}/C_3^{\rm Au}\approx 172\,$nm for the Au wall and
He${}^{\ast}$ atom.

In Fig.~2(a) the phenomenological interaction energy (\ref{eqn1})
multiplied by a factor $a^4$ is plotted as a function of
separation for the case of He${}^{\ast}$ atom interacting with Au
wall (the dashed line). In the same figure the solid line shows
the computational results for the quantity $a^4E(a)$, where the
accurate interaction energy $E(a)={\cal E}(a)$ is defined in
Eq.~(\ref{X6e20}) in accordance with the Lifshitz formula
at zero temperature. As is
seen in Fig.~2(a), at short and large separations the
phenomenological potential (\ref{eqn1}) almost coincides with the
accurate interaction energy, as given by the Lifshitz formula. To
give a better understanding of the correlation of the two
potentials at separations below $1\,\mu$m, i.e., in the most
important region for the experiments on quantum reflection, in
Fig.~2(b) both lines are shown in an enlarged scale. It is seen
that the solid and dashed lines coincide at $a\leq 50\,$nm. The
relative difference between the accurate and phenomenological
interaction energy,
\begin{equation}
\delta E(a)=\frac{E_{\rm acc}(a)-E_{\rm ph}(a)}{E_{\rm acc}(a)},
\label{eqn2}
\end{equation}
\noindent
is a nonmonotonous function and varies from 5.7\% at
$a=100\,$nm to 7.9\% at $a=1\,\mu$m. The largest values of $\delta E$
are achieved at moderate separations, which are interesting from
the experimental point of view: $\delta E=10.2$\%, 10.4\% and
10.2\% at separation distances $a=300$, 400 and 500\,nm,
respectively.

Now we consider the atom of metastable He${}^{\ast}$ near a high
resistivity Si wall (dielectric materials are often used in the
experiments on quantum reflection). In this case Eq.~(\ref{X6e27})
is not applicable. The value of the Casimir-Polder coefficient
$C_4^{\rm Si}\approx 0.75\,\mbox{eV\,nm}^4$ was computed in
Ref.~\cite{22} using the Lifshitz formula. The permittivity of
dielectric Si along the imaginary frequency axis with
$\varepsilon^{\rm Si}(0)=11.66$ was computed from the tabulated
optical data and Kramers-Kronig relations \cite{8,40}. In a
similar way the value of the van der Waals coefficient $C_3^{\rm
Si}\approx 5.5\times 10^{-3}\,\mbox{eV\,nm}^4$ was obtained in
Refs.~\cite{8,22}. This leads to $l^{\rm Si}\approx 136\,$nm for a
He${}^{\ast}$ atom near Si wall.

As an illustration, in Fig.~3(a) we plot the phenomenological
interaction energy (\ref{eqn1}) multiplied by a factor $a^4$ as a
function of separation for the case of He${}^{\ast}$ atom
interacting with Si wall (the dashed line). The solid line
presents the computational results for the quantity $a^4|E(a)|$
obtained using the Lifshitz formula as described above. In
Fig.~3(b) the same lines are reproduced on an enlarged scale at
separations below $1\,\mu$m. As is seen in Figs.~3(a,b), at
separations below 50\,nm and at about $10\,\mu$m the limiting
cases of a nonrelativistic and relativistic potentials $V_3$ and
$V_4$, respectively, are achieved. The relative difference
(\ref{eqn2}) between the accurate and phenomenological interaction
energies varies from 9.4\% at $a=100\,$nm to 8.6\% at $a=1\,\mu$m.
However, at intermediate separations $\delta E$ achieves the
largest values which are equal to 12\%, 12.5\%, 12.2\% and 11.6\%
at separations $a=200$, 300, 400, and 500\,nm, respectively. Thus,
for the Au wall the phenomenological interaction energy provides a
more accurate model of atom-wall interaction than for the Si wall.
This is connected with the fact that the strength of atom-wall
interaction for Si wall is weaker than in the case of an Au wall.

The above computations using the Lifshitz formula were performed
at zero temperature. It is instructive to compare the
phenomenological potential (\ref{eqn1}) with the results of more
accurate computations using the Lifshitz formula at the
temperature at laboratory $T=300\,$K. Computations were performed
by substituting the same, as above, dielectric permittivity of Au
and Si and dymanic polarizability of He${}^{\ast}$ atom along the
imaginary frequency axis into Eq.~(\ref{X6e17}). The computational
results for the quantity $a^4|{\cal F}(a,T)|$ are shown by the
solid lines in Fig.~4(a) for the Au wall and in Fig.~4(b) for the
Si wall. In addition, in these figures the same results, as in
Figs.~2(a) and 3(a), for the quantity $a^4|E(a)|$, where $E(a)$ is
the phenomenological potential (\ref{eqn1}) for Au and Si walls,
respectively, are reproduced with dashed lines. {}From the
comparison between Figs.~2(a) and 4(a) it is seen that for an Au
wall at separations $a\leq 2\,\mu$m from He${}^{\ast}$ atom the
relative differences between the accurate and phenomenological
potentials are approximately the same in the cases when the
accurate potential is computed at zero temperature or at
$T=300\,$K. However, with the increase of separation the accurate
potential, i.e., the free energy, computed at $T=300\,$K (the
solid line) deviates significantly from the phenomenological
potential in accordance with the classical limit in
Eq.~(\ref{X6e44}). For Au [Fig.~4(a)] the largest deviation shown
in the figure is equal to 31\%. It is achieved at $a=5\,\mu$m.

For the He${}^{\ast}$ atom near the Si wall [Fig.~4(b)], the thermal
effects play a more important role. The comparison between  Figs.~3(a)
and 4(b) demonstrates that here the differences between the accurate,
${\cal F}(a,T)$, and phenomenological, $E(a)$, potentials can be
considered as temperature-independent only below $0.5\,\mu$m.
Computations show that at $a=1\,\mu$m the relative difference between
them is equal to 9.6\% (whereas, as indicated above, it is equal to only
8.6\% when the zero-temperature Lifshitz formula is used).
At the separation $a=5\,\mu$m the relative difference between the accurate
temperature-dependent and phenomenological potentials achieves 43.5\%.

Larger deviations between the accurate temperature-dependent
potential and the phenomenological potential for dielectrics than
for metals are explained by the existence of temperature and
separation regions where the Casimir-Polder entropy is negative
(see Secs.~III--V). The phenomenon of negative entropy occurs only
for atoms near a metallic plate. As a result, within some range of
temperatures, the sign of the thermal correction to the
Casimir-Polder energy is opposite to the sign of the energy and
the respective free energy becomes nonmonotonic. This makes
smaller the difference between the accurate free energy, as
computed by the Lifshitz formula, and the phenomenological
potential. On the contrary, for an atom near a dielectric wall the
Casimir-Polder entropy is always positive \cite{41}. This follows
from the same property of the entropy in the configuration of two
dielectric plates \cite{24,46}. Then the thermal correction and
the Casimir-Polder energy have the same sign and the magnitude of
the free energy is a monotonously increasing function of the
temperature. Thus, with the increase of temperature (or
separation) differences between the accurate free energy and the
phenomenological potential can only increase.

\section{Conclusions and discussion}

In the foregoing we have investigated some novel aspects of the
Casimir-Polder interaction between an atom and a metal wall. For
an ideal metal wall, the cases of short separations on the one
hand, and moderate and large separations, on the other hand, were
considered. At short separations, the delicate problem connected
with the dependence of the Casimir-Polder free energy on the
velocity of light was discussed. At moderate and large separations
the analytical expressions for the Casimir-Polder free energy,
force and entropy were obtained. It is shown that the Nernst heat
theorem is satisfied, but at low temperatures the Casimir-Polder
entropy takes negative values. This conclusion was extended to the
case of real metals. First, the role of different corrections due
to nonzero skin depth, dynamic atomic polarizability and nonzero
temperature was discussed and simple analytical expressions
applicable to real metal walls were obtained. Then, the asymptotic
behavior of the Casimir-Polder free energy and force was derived
for an atom near a real metal wall. This permitted us to prove the
Nernst heat theorem and demonstrates that the Casimir-Polder
entropy takes negative values within some temperature range, as is
the case of ideal metal wall.

The results obtained were applied to compare the phenomenological
potential used in the theoretical interpretations of experiments
on quantum reflection and the accurate atom-wall interaction
energy computed on the basis of the Lifshitz theory. This
comparison was performed for metastable He${}^{\ast}$ atoms within
a wide separation region from 20\,nm to $10\,\mu$m for both metal
and dielectric walls. It was shown that at separations below
$1\,\mu$m the phenomenological potential deviates from the
accurate one up to 10.4\% for a metal (Au) wall and up to 12.5\%
for a dielectric (Si) wall. At a separation $a=5\,\mu$m, where the
thermal effect plays an important role, the relative differences
between the accurate and phenomenological potentials achieve 31\%
and 43.5\% for metallic and dielectric walls, respectively. The
decreased relative differences of the accurate and
phenomenological potentials for metal walls are explained by the
negativeness of the Casimir-Polder entropy. Bearing in mind that
most of experiments on quantum reflection are performed with
dielectric walls, the use of a more accurate potential seems to be
preferable for the comparison of the measurement data with theory.

\section*{Acknowledgments}

V.B.B. and C.R. are grateful to CNPq, FAPESQ-PB/CNPq (PRONEX) and
FAPES-ES/CNPq (PRONEX) for partial financial support. G.L.K.\ and
V.M.M.\ were supported by Deutsche Forschungsgemeinschaft,
 Grant No 436 RUS 113/789/0--4.

\begin{figure*}[h]
\vspace*{-16.7cm}
\centerline{
\includegraphics{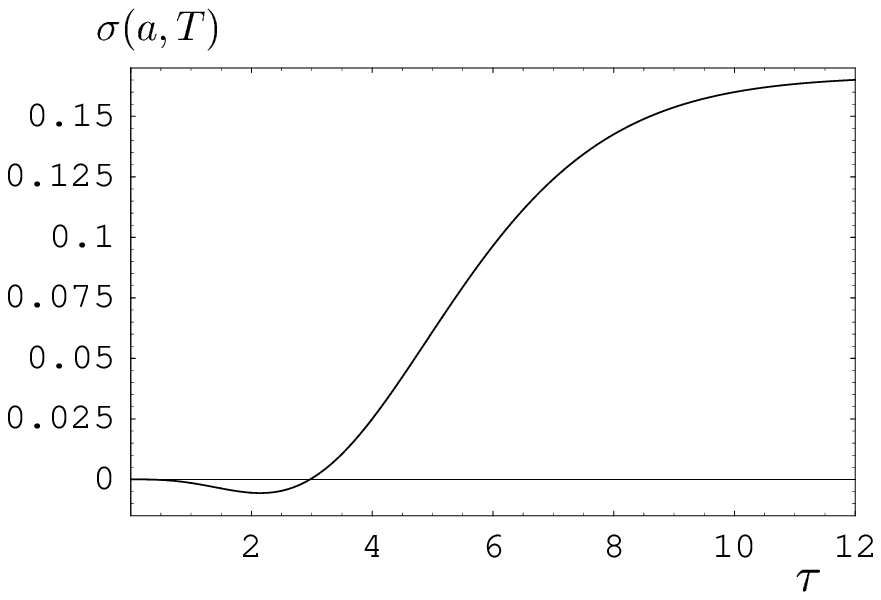}
}
\vspace*{-7.3cm}
\caption{The entropy factor $\sigma$ from Eq.~(\ref{X6e42})
for an atom near an ideal metal wall as a function
of $\tau$.}
\end{figure*}
\begin{figure*}[h]
\vspace*{2.cm}
\centerline{
\includegraphics{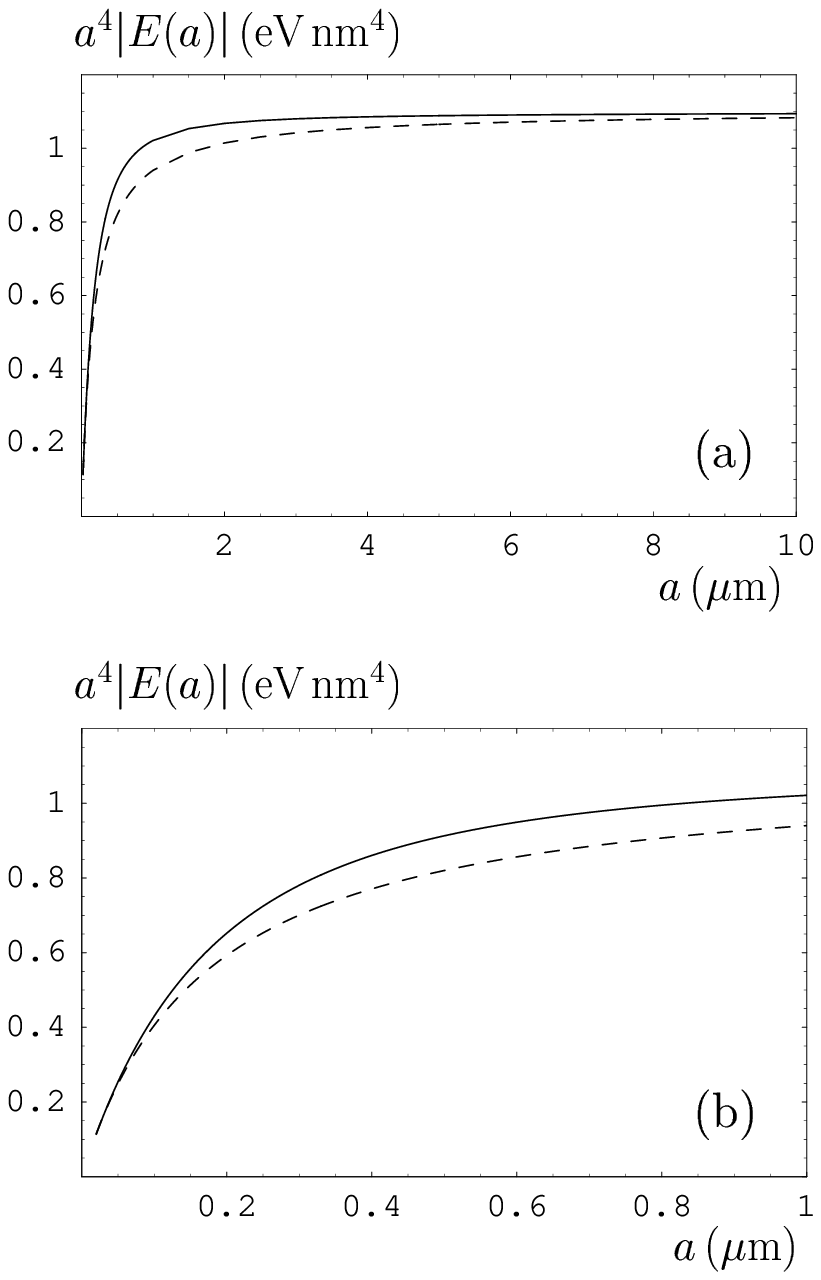}
}
\vspace*{-16.3cm}
\caption{Magnitude of the interaction energy between an atom of
metastable He${}^{\ast}$ and Au wall multiplied by the fourth
power of separation versus separation. Computations are performed
using the Lifshitz formula at $T=0$ (the solid lines) and the
phenomenological potential (\ref{eqn1}) (the dashed lines).
(a) Separation varies from 20\,nm to $10\,\mu$m.
(b) Separation varies from 20\,nm to $1\,\mu$m.}
\end{figure*}
\begin{figure*}[h]
\vspace*{2.cm}
\centerline{
\includegraphics{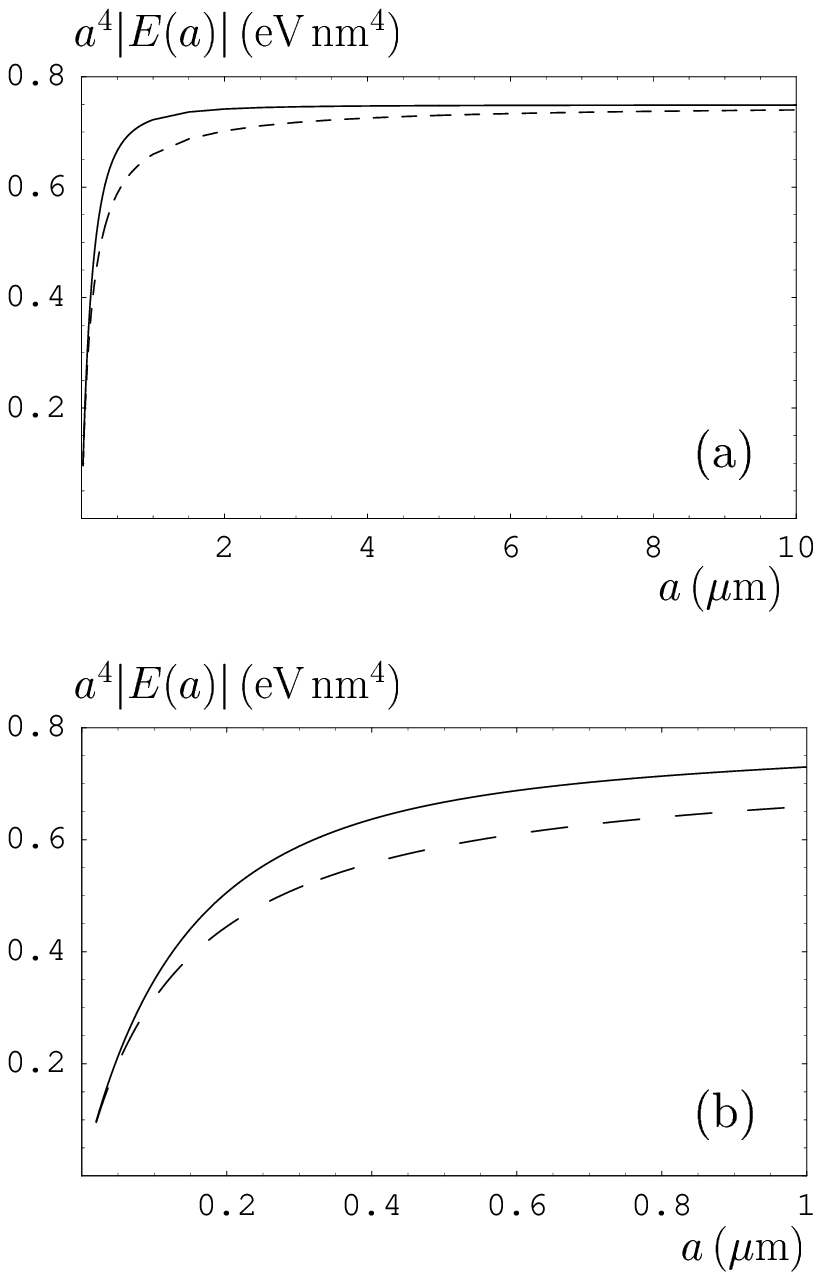}
}
\vspace*{-16.3cm}
\caption{Magnitude of the interaction energy between an atom of
metastable He${}^{\ast}$ and Si wall multiplied by the fourth
power of separation versus separation. Computations are performed
using the Lifshitz formula at $T=0$ (the solid lines) and the
phenomenological potential (\ref{eqn1}) (the dashed lines).
(a) Separation varies from 20\,nm to $10\,\mu$m.
(b) Separation varies from 20\,nm to $1\,\mu$m.}
\end{figure*}
\begin{figure*}[h]
\vspace*{2.cm}
\centerline{
\includegraphics{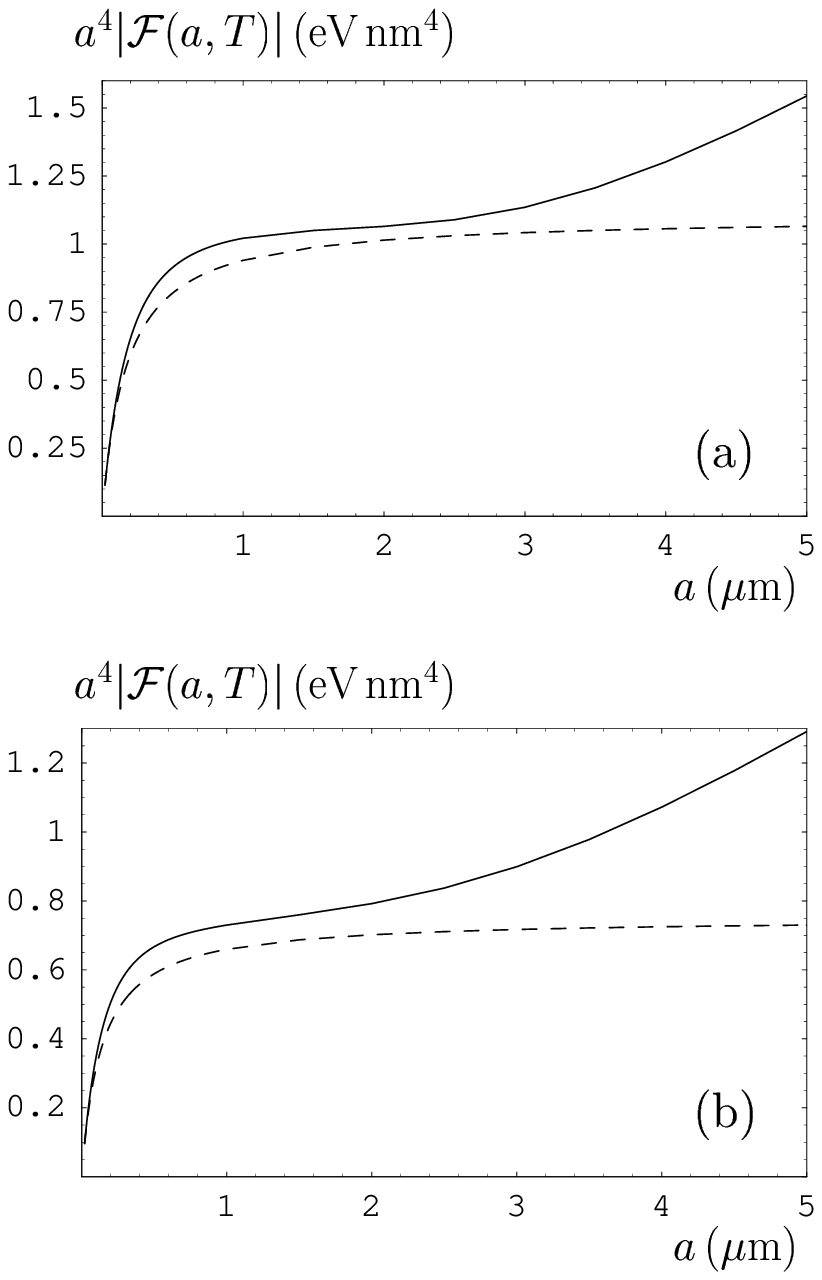}
}
\vspace*{-16.3cm}
\caption{Magnitude of the free energy between an atom of
metastable He${}^{\ast}$ and (a) Au or (b) Si wall multiplied
by the fourth power of separation  as a function
of separation is shown by the solid lines.
The free energy is computed at $T=300\,$K using the
Lifshitz theory. For comparison the phenomenological potential (\ref{eqn1})
for (a) Au and (b) Si walls is shown by the dashed lines.}
\end{figure*}
\end{document}